\documentclass[12pt]{article}
\usepackage{geometry}                
\usepackage{graphicx}
\usepackage{amssymb}
\usepackage{epstopdf}
\def\hk{ H^{(k)}}

\def\call{{\cal L}}
\def\callk{{\cal L}^{(k)}}
\def\calg{{\cal G}}

\def\b2hat{ {\hat b}_2 }

\def\bhat{\hat b}

\def\kmax{k_{max}}
\def\calh{{\cal H}}

\begin{document}

\begin{titlepage}
\vfill
\begin{flushright}
\end{flushright}
\vskip 1.0in

\begin{center}
\baselineskip=16pt

{\Large\bf Mass and Free Energy of Lovelock Black Holes}
\vskip 0.5cm
{\large {\sl }}
\vskip 5.mm
{\bf David Kastor${}^{a1}$, Sourya Ray${}^{b2}$ and Jennie Traschen${}^{a3}$} \\

\vskip 0.5cm
{
	${}^a$ Department of Physics, University of Massachusetts, Amherst, MA 01003\\	
     	${}^b$ Centro de Estudios Cientõficos (CECS), Casilla 1469, Valdivia, Chile \\
	${}^1$\texttt{kastor@physics.umass.edu,} ${}^2$ \texttt{ray@cecs.cl,} ${}^3$ \texttt{traschen@physics.umass.edu}

     }
\vspace{6pt}
\end{center}
\vskip 0.5in
\par
\begin{center}
{\bf Abstract}
 \end{center}
\begin{quote}
An explicit formula for the ADM mass of an asymptotically AdS black hole in a generic Lovelock gravity theory is presented, identical in form to that in Einstein gravity, but  multiplied by a function of the Lovelock coupling constants and the AdS curvature radius.
A Gauss' law type formula relates the mass, which is an  integral at infinity, to an expression depending instead on the horizon radius.  This and other thermodynamic quantities, such as the free energy, are then analyzed in the limits of small and large horizon radius, yielding results that are independent of the detailed choice of Lovelock couplings. In even dimensions, the temperature diverges in both limits, implying the existence of a minimum temperature for black holes.  The negative free energy of sufficiently large black holes implies the existence of a Hawking-Page transition.
In odd dimensions the temperature still diverges for large black holes, which again have negative free energy.  However, the temperature vanishes as the horizon radius tends to zero and sufficiently small black holes have positive specific heat.
\vskip 2.mm
\end{quote}
\end{titlepage}


\section{Introduction}

Many interesting, stationary black holes are not known in analytic form.  The list includes charged rotating black holes in $D>4$, black rings in $D>5$, localized Kaluza-Klein black holes and rotating black holes in Lovelock gravity theories.  
Results have been obtained using a variety of approximate techniques, including {\it e.g.} perturbative expansions in the slowly rotating limit \cite{Aliev:2005npa,Aliev:2006yk,Kim:2007iw,Zou:2010dx,Yue:2011et}  or effective field theory methods \cite{Emparan:2009cs,Emparan:2009at,Caldarelli:2010xz}.  
However, one may also ask whether in the absence of analytic solutions,  it might still be possible to at least obtain certain properties of such spacetimes, such as thermodynamic ones, exactly.  

In this paper we will address this question for a class of spacetimes that might best be called ``semi-unknown", namely static Lovelock black holes.  It is known \cite{Wheeler:1985qd} that these spacetimes are Schwarzschild-like, in the sense that the metric written in the  general static, spherically symmetric form
\begin{equation}\label{generalmetric}
ds^2 = - \phi(r) dt^2 + {dr^2\over f(r)} +r^2d\Omega_{D-2}^2,
\end{equation}
additionally has $\phi(r)=f(r)$.  For a generic Lovelock theory, the function $f(r)$ must satisfy an algebraic equation of order 
$\left[ {D-1\over 2}\right]$, where $D$ is the spacetime dimension and the closed brackets denote taking the integer part.  The coefficients of the polynomial are  the coupling constants of the higher curvature terms in the Lovelock Lagrangian.  Since the roots of a generic polynomial equation can be found in terms of radicals only up to order $4$, it follows that in spacetimes dimensions $D>10$ the general solution for a static Lovelock black hole cannot be written down in a simple closed form and hence remains unknown\footnote{Given that the solutions to generic cubic and quartic equations are quite cumbersome, in practice the general solution is only known in useful form for $D\le 6$, although certain non-generic solutions such as those for Gauss-Bonnet gravity \cite{Boulware:1985wk,Wheeler:1985nh,Cai:2001dz,Cai:2003gr}, or Chern-Simons gravity \cite{Cai:1998vy,Crisostomo:2000bb} are known in all dimensions.  Note also that an analytic solution for a rotating black hole in $D=5$ Gauss-Bonnet gravity has also been found in the case of Chern-Simons couplings \cite{Anabalon:2009kq} (see also \cite{Anabalon:2010ns}).}.

We make use of results from references \cite{Kastor:2008xb,Kastor:2009wy,Kastor:2010gq},  derived using the  Hamiltonian formulation \cite{lovelock-hamiltonian}, to obtain exact properties of Lovelock black holes, without requiring the explicit (and in general unknown) form of the metric function $f(r)$. In fact with future applications to other, even more unknown, black holes in mind, we will ``forget" that we know that $\phi(r)=f(r)$ and take the general static, spherically symmetric form (\ref{generalmetric}) as our starting point.  Our focus throughout will be on black holes with asymptotically AdS boundary conditions.

Further motivation comes from recent work on the CFT duals of asymptotically AdS solutions to higher curvature gravities. Significant work has been
done, for example, on the relation between CFT plasmas and their gravitational duals,
\cite{Dias:2009iu,Caldarelli:2008mv,Aharony:2005bm,Lahiri:2007ae} 
  and on implications for the CFT of causality and stability in the bulk using  holography
\cite{Kovtun:2004de,Kats:2007mq,Brigante:2008gz,Ge:2008ni,Ge:2009eh,Ge:2009ac,Buchel:2009sk,Buchel:2009tt,deBoer:2009gx,Hofman:2009ug,Shu:2009ax,deBoer:2009pn,Camanho:2009vw,Camanho:2009hu,Ge:2010aa}.
As the correspondence between the  higher curvature bulk theory and the boundary CFT becomes better understood,  
it is interesting to ask whether the constructs we will make use of below, the Killing-Lovelock potentials and the associated Gauss' law relations that connect behavior at the horizon to behavior at infinity, have an analogue in the dual CFT.

The paper will proceed as follows.  In section (\ref{lovelocksection}) we  recall the basic elements of Lovelock gravity theories  \cite{Lovelock:1971yv} that will subsequently be used.
  In section (\ref{masssection}) we will derive the expression for the ADM mass of an asymptotically AdS Lovelock black hole in terms of its far field behavior.  
  The ADM boundary integral receives
contributions from each higher curvature term as well as from the Einstein term.
  The formula for the ADM mass is implicit in the work of \cite{Jacobson:1993xs} which computed the entropy in Lovelock theories and established the first law.    However significant additional steps are required to make the result of \cite{Jacobson:1993xs} fully explicit.   The  expression we find ultimately takes the simple form of the standard ADM mass integral multiplied by a function of the Lovelock couplings and the asymptotic AdS curvature radius.  This formula for the mass constitutes a new
  result.

In section (\ref{kl_potentials}) we review certain elements from our previous work \cite{Kastor:2008xb,Kastor:2009wy,Kastor:2010gq}
that we will be making use of.  This includes the Killing-Lovelock potentials which allow for the derivation of a Gauss' law type expression for the Hamiltonian constraint, which will be our main tool.  This expression relates certain boundary integrals at infinity to surface integrals on the horizon, without requiring the explicit form of the metric in between.   For reference, we also present the Smarr formula  \cite{Kastor:2010gq} which gives the mass in terms of the entropy and additional thermodynamic quantities that arise in the Lovelock theory.  

The bulk of our new results appear in sections (\ref{mainsection}) and (\ref{freesection}).  In section (\ref{mainsection}) we use the Gauss' law formula to obtain an expression for the mass, originally given by the far field behavior of the metric, in terms of the horizon radius of the black hole.  With the goal in mind of providing a simple expression for the free energy, we also present formulas for the entropy, surface gravity and other thermodynamic quantities in terms of the horizon radius.  The detailed behavior of these expressions over the full range of horizon radii depends on the choice of Lovelock coupling constants.  However, we find that they have generic behavior in the limits of small and large horizon 
radii. There are qualitative differences between even and odd dimensions that
arise in the small black hole regime.  In particular, one finds that in odd dimensions there exists a ``mass gap", in the sense that the mass tends to a finite value
 as the horizon radius tends to zero.  The Hawking temperature also vanishes in this limit and sufficiently small black holes have positive specific heat.  For even dimensions, however, the mass tends to zero for vanishing horizon radius and the specific heat for small black holes is negative.

In section (\ref{freesection}) we assemble ingredients to construct a relatively simple expression for the free energy of black holes in generic Lovelock theories.  The free energy had previously been computed only up to inclusion of the quadratic Gauss-Bonnet term \cite{Cai:2001dz,Cvetic:2001bk} in the Lovelock Lagrangian.  The free energy for a stationary black hole solution is generally obtained by computing its Euclidean action. We emphasize that our method does not require the explicit functional form of the metric.  We analyze the free energy and also the specific heat in the small and large black hole limits and comment upon the Hawking-Page phase transition for generic AdS-Lovelock black holes.  In section (\ref{discussion}) we briefly summarize our results and offer some directions for further related work.

\section{Lovelock gravity}\label{lovelocksection}
The Lagrangian of a Lovelock gravity theory  in $D$ spacetime dimensions 
is given by 
\begin{equation}\label{lovelocklagrangian}
{\cal L}={1\over 16 \pi G} \sum_{k= 0}^{\kmax}b_k{\cal L}^{(k)}
\end{equation}
where $\kmax = [(D-1)/2]$ and the $b_k$ are real-valued coupling constants.  The symbol
$\callk$ stands for the contraction of $k$ powers of the Riemann tensor given by
\begin{equation}\label{lovelagran}
\call^{(k)} ={1\over 2^k } \delta ^{a_1 b_1...a_k b_k } _{c_1 d_1 ....c_k d_k }
 R_{a_1 b_1}{}^{c_1 d_1 }\dots  R_{a_k b_k}{}^{c_k d_k }.
\end{equation}
where the $\delta$-symbol is the totally anti-symmetrized product
normalized so that it takes nonzero values $\pm 1$.  
The term $\call^{(0)}$ gives the cosmological constant term in the Lagrangian, while $\call^{(1)}$ gives the Einstein-Hilbert term and $\call^{(2)}$ the quadratic Gauss-Bonnet term. The upper bound in the sum (\ref{lovelocklagrangian}) comes about because $\call^{(k)}$ vanishes identically for $D<2k$ and turns out to make no contribution to the equations of motion in $D=2k$.  

The equations of motion for Lovelock gravity can be written as $\calg^a{}_b= 0$ where 
\begin{equation}
\calg^a{}_b= \sum_{k=0}^{\kmax} b_k{\cal G}^{(k)a}{}_b. 
\end{equation}
We will not need the explicit expressions for these quantities.  However, 
it is crucial in what follows that  each of the tensors in this sum satisfies a conservation law $\nabla _a {\cal G}^{(k)a}{} _b  =0$.  
 
Depending on the values of the coupling constants $b_k$ the theory may have anywhere from zero up to $\kmax$ distinct  constant curvature vacuum solutions.   Because we will be focusing on asymptotically AdS black holes below, our considerations will implicitly be limited to the subset of theories admitting at least one constant negative curvature vacuum.  We will denote the curvature radius of this AdS vacuum by $l$.

We will also need the Hamiltonian formulation of  Lovelock gravity which was developed in \cite{lovelock-hamiltonian}.
As usual in the Hamiltonian picture the spacetime metric is split according to 
 \begin{equation}\label{metricsplit}
g_{ab}=-n_a n_b +s_{ab}
\end{equation}
where $n^a$ is the unit timelike normal to a spatial slice $\Sigma$ with induced metric $s_{ab}$ and these satisfy the orthogonality relation $s_{ab} n^b =0$.  As in Einstein gravity the time-time and time-space components of the field equations act as constraints on initial data.
In Lovelock gravity the Hamiltonian and momentum constraint operators $H = -2n^an^b\,\calg_{ab}$ and $H_a = -2s_a{}^b\,n^c\, \calg_{bc}$
are given by the sums
\begin{equation}
H= \sum_{k=0}^{\kmax} b_k\, H^{(k)},\qquad H_a= \sum_{k=0}^{\kmax} b_k\, H_a^{(k)}
\end{equation}
with $H^{(k)}=-2n^an^b\,\calg^{(k)}_{ab}$ and $H_a^{(k)}=-2s_a{}^b\,n^c\, \calg^{(k)}_{bc}$.
We will particularly need the expression for
\begin{equation}\label{hperp}
H^{(k)} = -\, {1\over 2^k}\, \tilde\delta ^{a_1 b_1...a_k b_k } _{c_1 d_1 ....c_k d_k }\, 
 R_{a_1 b_1}{}^{c_1 d_1 }\dots  R_{a_k b_k}{}^{c_k d_k }
\end{equation}
where the tilde on the $\delta$-symbol indicates that its indices are projected with the spatial metric $s^a{}_b$.
The actual Hamiltonian function for evolution of initial data with respect to a vector field $\xi^a$  is then given by 
${\cal H}_\xi =F\,H +F^a\,H_a$, 
where the lapse and shift $(F,F^a)$ are the components of the vector field $\xi^a$ normal to and along the  spatial slice, so that $\xi ^a = Fn^a +F ^a$.  As in Einstein gravity  the lapse  and shift are Lagrange multipliers.

Finally, for use below we define the sums 
\begin{equation}\label{sums}
s_{(n)} = \sum_{k=0}^{\kmax}(-1)^{k+n}{k!\,\,\bhat_k\over (k-n)!\,\,  l^{2k-2}}
\end{equation}
where the rescaled coefficients $\bhat_k$ are given by $\bhat_k = (D-1)!\, b_k/(D-2k-1)!$.
Note that the combinations $\bhat_k/l^{2(k-1)}$ are dimensionless.  The condition $s_{(0)}=0$, in fact, determines the allowed constant curvature vacua of the theory, while the sums $s_{(1)}$ and $s_{(2)}$ will turn up in our results below.

\section{ADM Mass}\label{masssection}

An expression for the ADM mass of an asymptotically AdS Lovelock black hole can be obtained using the methods of Regge and Teitelboim 
\cite{Regge:1974zd}. We find a simple, explicit formula which is similar to that for the ADM mass.
 As in Einstein gravity,  a boundary term must be included in the Lovelock Hamiltonian to ensure that the Hamiltonian variational principle correctly yields the equations of motion.  The variation of this boundary term cancels another boundary term arising via integration by parts from the variation of the bulk Hamiltonian.  The ADM mass is defined to be the value of the Hamiltonian divided by a factor of $16 \pi G$.  Since the bulk Hamiltonian vanishes on solutions, the ADM mass is simply proportional to the value of the Hamiltonian boundary term.  Implicit here is that the vector $\xi^a$ determining the direction of Hamiltonian evolution should be asymptotic to the time translation Killing vector of AdS.

This procedure was formally carried out in reference \cite{Jacobson:1993xs} as part of establishing the first law for Lovelock black holes.  Each higher curvature term in the Lovelock theory makes a contribution to the Hamiltonian boundary term.  For asymptotically flat solutions, possible if $b_0=0$ in the Lagrangian (\ref{lovelocklagrangian}), because of the fall-off of the curvature tensor, only the boundary term corresponding to the Einstein term is nonzero at infinity.  The formula for the mass is then the same as in Einstein gravity.  However,  the non-zero asymptotic value of the curvature tensor for AdS Lovelock black holes leads to contributions from all the higher curvature boundary terms.   We will see, in fact, that the result is given by the usual ADM integral multiplied by a function of the Lovelock couplings and the background AdS curvature. To our knowledge this expression has not been derived in the literature\footnote{A number of related results have been established in the literature.  The Hamiltonian approach of \cite{Regge:1974zd} has been used to obtain an expression for the mass
in pure Lovelock gravity theories \cite{Crisostomo:2000bb} with only a single
term in Lagrangian.  Significant work has also been done in the case of Gauss-Bonnet gravity.
Deser and Tekin have extended the formalism of Abbott and Deser  \cite{Abbott:1981ff} to general quadratic theories of gravity \cite{Deser:2002rt,Deser:2002jk}.   
The mass in Gauss-Bonnet gravity has also been found by means of a Noether's current construction by Deruelle et. al. \cite{Deruelle:2003ps} and 
using the Palatini formalism by Katz and Livshits \cite{katz}  The mass of asymptotically AdS black holes in general higher curvature theories has also been addressed in \cite{Okuyama:2005fg}.

Padilla \cite{Padilla:2003qi} uses the Cartan formalism to derive a general expression for the mass in Gauss-Bonnet gravity by transforming the trace $K$ boundary term in the action into Hamiltonian variables.   By assuming that a generalization of this result holds in general Lovelock theories, the mass of static Lovelock black hole is identified by Cai \cite{Cai:2003kt}  with a constant of integration arising
from integration of the Hamiltonian constraint.   Our general derivation of the mass in Lovelock theories demonstrates the correctness of this assumption.}.

The Regge-Teitelboim type construction of the ADM mass in  \cite{Jacobson:1993xs} yields an expression of the form
\begin{equation}\label{lovelockmass}
M = - {1\over 16\pi G}\sum_{k=0}^{\kmax} b_k\int_\infty da_c B^{(k)c}.
\end{equation}
We will determine the boundary integrands $B^{(k)c}$ explicitly for asymptotically AdS boundary conditions.  
Because our interest is focused on static black holes, we will assume that the momentum  
and the shift vector 
vanish sufficiently rapidly in the asymptotic  region that they do not play a role in this construction.  It is then sufficient to retain only the term  
$\calh_\xi=F H$ in the Hamiltonian and to take the curvature tensor in (\ref{hperp}) to be that of the spatial metric $s_{ab}$.  
Now, assume that the spatial metric $s_{ab}$ solves the Lovelock constraint equations with asymptotically AdS boundary conditions and add to it an arbitrary perturbation $h_{ab}$ that also respects these boundary conditions.
To first order, the perturbation to the Riemann tensor is then $\delta R_{ab}^{\ \ cd} = {R}_{ab}^{\ \ e[c}h_e ^{\  d]} -2 D _{[a}D ^{[c} h_{b]} ^{\ \  d]}$, where $D_a$ is the covariant derivative operator for the spatial metric $s_{ab}$.
Plugging this in to the variation of (\ref{hperp})  yields the perturbation of the $k$th  Hamiltonian
\begin{equation}\label{deltahk}
\delta H^{(k)} =  -{k\over 2^k}\, \tilde\delta ^{a_1 b_1...a_k b_k } _{c_1 d_1 ....c_k d_k }\, 
 {R}_{a_1 b_1}{}^{c_1 d_1 }\dots  {R}_{a_{k-1} b_{k-1}}{}^{c_{k-1} d_{k-1} }
\left(  {R}_{a_k b_k }{}^{ e c_k  }h_e{}^{d_k} -2 D _{a_k} D ^{c_k} h_{b_k}{}^{d_k} \right) 
\end{equation}
Combining these, multiplying by the lapse function $F$, integrating by parts, and making use of the Bianchi identity for the Riemann tensor 
then gives an overall expression of the form 
\begin{equation}
F\delta H = \delta H^* \cdot F +\sum_k b_k D_c B^{(k)c}.
\end{equation}
Here the first term represents an adjoint differential operator acting on the lapse function $F$ and the vectors $B^{(k)c}$ are given by
\begin{equation}\label{btk}
 B^{(k)c} =   {k\over 2^k}\, \tilde\delta ^{cd m_1 n_1...m_k n_k } _{ab e_1 f_1 ....e_k f_k }\, 
 {R}_{m_1 n_1}{}^{e_1 f_1 }\dots  {R}_{m_k n_{k-1}}{}^{e_{k-1} f_{k-1} }
 \left( F D^a h_d {}^b  -h_d{} ^b D^a F \right)
 \end{equation}

These are the vectors  appearing in the expression for the ADM mass given above.  The quantity $h_{ab}$ is then the deviation of the static black hole metric from the asymptotic AdS background, while the minus sign in  (\ref{lovelockmass}) arises from the cancellation of boundary terms required in the Regge-Teitelboim construction.
The expression can be made more fully explicit by noting that near infinity only the leading order background AdS curvature  $R_{ab }{}^{cd}
=(1/ l^2) \delta _{ab} ^{cd}$ contributes. 
This leads to the result
\begin{equation}\label{finalmass}
M =
{s_{(1)}\over  (D-1)(D-2)}\cdot \left( {-1\over 16\pi G}
\int_\infty da_c\left\{ F(D^ch-D_bh^{cb}) - hD^cF +h^{cb}D_b F\right\}\right).
\end{equation}
The part of the formula in parenthesis is the usual expression for the ADM mass.  The sum  $s_{(1)}$, defined in (\ref{sums}), depends on the Lovelock coupling constants and the curvature radius $l$ of the asymptotic AdS vacuum.  For Einstein gravity with a cosmological constant  the prefactor outside the parenthesis reduces to one.  The asymptotic form of the lapse function is $F=\sqrt{1+ r^2 / l^2}$, and therefore terms in (\ref{finalmass}) involving the derivative of the lapse can make non-trivial contributions to the mass. In the asymptotically
flat case, these terms fall off too quickly to contribute and the integrand reduces to that of the ADM mass in the asymptotically flat case.  

\section{Killing-Lovelock Potentials}\label{kl_potentials}

Killing-Lovelock potentials \cite{Kastor:2008xb,Kastor:2010gq} allow the Lovelock Hamiltonian constraint equations for a spacetime with a Killing vector to be written in a Gauss' law form.  In reference \cite{Kastor:2010gq} this property was used to derive an extended first law including variation of the Lovelock coupling constants and also a related Smarr formula for Lovelock black holes.  This will also be our primary tool below.

For a spacetime, such as a stationary black hole, with a Killing vector $\xi^a$ the Killing-Lovelock potentials were defined in \cite{Kastor:2008xb} to be antisymmetric tensors  $ \beta^{(k)ab}$ satisfying the relations
\begin{equation}\label{potdef}
\nabla_a \beta^{(k)ab} = -2  {\cal G}^{(k)a}{} _b\xi^b.
\end{equation}
Their existence is guaranteed by the vanishing divergence of the right hand side.  However,
they are not uniquely determined, since a divergenceless tensor such as the divergence of an arbitrary $3$-index antisymmetric tensor may always be added.  This ambiguity does not affect results such as the extended first law or Smarr formula.

Now consider Hamiltonian evolution with respect to the Killing vector, so that the lapse and shift are given by the decomposition $\xi ^a =Fn^a +F^a$.
The definition of the Killing-Lovelock potentials allows us to write the $k$th Hamiltonian function as a total divergence, 
\begin{equation}\label{kthhamdiv}
\calh^{(k)}=  F \hk +F^a \hk _a 
= -2 {\cal G}^{(k)d}  _c \xi ^c n_d = D_c ( \beta^{(k)cd} n_d ).
\end{equation}  
The Hamiltonian constraint equation $\calh=0$ can then be written in the form of Gauss's law.  
It ultimately follows from this (see \cite{Kastor:2010gq} for how this works) that the Killing-Lovelock potentials make contributions to the thermodynamics  of AdS-Lovelock black holes that are similar in form to that of the electrostatic potential in the case of a charged black hole.   For example, one contribution to the change in 
energy in the extended first law of \cite{Kastor:2010gq} is proportional to the difference in the integrals of Killing potential over the boundaries of a spatial slice at infinity and at the black hole horizon.  The AdS boundary conditions require a subtraction at infinity of  the Killing-Lovelock potential  
$\beta _{AdS} ^{(k)ab}$ for the asymptotic AdS spacetime resulting in
thermodynamic potentials  
\begin{equation}\label{theta}
\Theta ^{(k)} =- \left( \int _\infty  da r_c  (\beta^{(k)cd}
 - \beta _{AdS} ^{(k)cd} )n_d - \int _{h}  da r_c \beta^{(k)cd} n_d \right).
\end{equation}
that multiply variations $\delta b_k$ of the Lovelock couplings in the extended first law.
As a consequence of the constraint  equations these thermodynamic potentials satisfy the sum rule
\begin{equation}\label{potsum}
\sum _{k=0}^{\kmax} b_k \Theta ^{(k)} =0.
\end{equation}

The Smarr formula relates the mass of a black hole which is defined in terms of the behavior of the metric near infinity to the entropy which comes from the behavior at the horizon.  For static Lovelock black holes the Smarr formula \cite{Kastor:2010gq}, which we will also make use of below, is given by
\begin{equation}\label{smarr}
\left( (D-3) +2{s_{(2)} \over s_{(1)} }\right)  M ={\kappa \over 2\pi} \left[  (D-2) S - S^\prime\right]
  -\Theta
\end{equation}
The  entropy $S$  \cite{Jacobson:1993xs} has contributions from the higher curvature Lovelock terms and is given by 
$S=\hat A/4G$ with $\hat A=\sum_k b_k A_k$ and
\begin{equation}
A_k =k\,  \int_{h} d^{D-2}x \sqrt{\gamma}\, \call^{(k-1)}(\gamma_{ab})
\end{equation}
where $\gamma_{ab}$ is the induced metric on the boundary of the spatial slice at the horizon.
There is also an additional entropy-like contribution at the horizon as well as the net contribution Lovelock thermodynamic potentials which are
given respectively by
\begin{equation}\label{extras}
S^\prime={1\over 2G}\sum _{k=0}^{\kmax} (k-1)b_k  A^{(k)} , \qquad \Theta \equiv {1\over 8\pi G}\sum_{k=0}^{\kmax}  (k-1) b_k \Theta^{(k)}.
\end{equation}
The Smarr formula is derived in  \cite{Kastor:2010gq} via an overall scaling argument from an extended form of the first law in which variations of the dimensionful Lovelock coupling constants are taken into account.  
The contribution of $S^\prime$ to the Smarr formula arises from the explicit dependence of the entropy on the Lovelock couplings, and the second term in the parenthesis on the left in (\ref{smarr}) similarly arises from the dependence of the mass (\ref{lovelockmass}) on the couplings $b_k$.

 For Einstein gravity with vanishing cosmological constant one finds that $s_{(2)} =S^\prime=\Theta=0$ and one recovers Smarr formula for static, asymptotically flat black holes.  The case of Einstein gravity with non-zero cosmological constant was discussed in 
 \cite{Kastor:2009wy}.  In this case, one still has $s_{(2)} =S^\prime=0$.  However, there is a non-trivial contribution to $\Theta$ from the product\footnote{One has $\Lambda=-2b_0$ in this case.}
 $\Lambda\Theta^{(0)}$.  In  \cite{Kastor:2009wy} it was argued that $\Theta^{(0)}$ can be regarded as minus an effective volume behind the black hole horizon.  Since the cosmological
 constant is proportional to minus the background pressure, 
 one sees that
 the $k=0$ term in the Smarr formula has the form  $pV$ which is familiar from
 classical thermodynamics.  The mass $M$ should then be regarded as the spacetime enthalpy.  This interpretation was explored further in \cite{Dolan:2010ha,Cvetic:2010jb}.

\section{Lovelock black holes}\label{mainsection}

In this section we will derive formulas for the basic thermodynamic properties of 
static, spherically symmetric AdS Lovelock black holes in terms of the horizon radius $r_h$ and the Lovelock parameters $b_k$.  
We consider  static, spherical symmetric, asymptotically AdS spacetimes with AdS curvature radius $l$.  The metric can then be taken to have the form
(\ref{generalmetric}), 
where at large radius the metric functions have the asymptotic forms
 \begin{equation}\label{falloff}
 \phi (r)\sim 1+ { r^2\over l^2} -{c_t \over r^{D-3}}\  , \quad  f (r)\sim 1+ { r^2\over l^2} -{c_r\over r^{D-3}}
 \end{equation}
for some constants $c_t$ and $c_r$. 
We assume that there is a Killing horizon at $r=r_h$, where $\phi (r_h ) =0$. 
Static spherically symmetric black hole solutions of Lovelock gravity theories have been known for some time, starting with the work of  \cite{Boulware:1985wk,Wheeler:1985nh} in the Gauss-Bonnet case and \cite{Wheeler:1985qd} in the general Lovelock case (see also \cite{Myers:1988ze}).  
These solutions all have $\phi(r)=f(r)$ and the field equations reduce to the requirement that $f(r)$ solve a certain polynomial equation of order 
$\kmax$, with coefficients determined by the Lovelock coupling constants\footnote{Specifically, if one defines the function $F$ according to 
$f=1-r^2 F$, then $F$ must satisfy the polynomial equation $ \sum_{k=0}^{\kmax} {\hat b}_k F^k = {\omega\over r^{D-1}}$
for some constant  $\omega$. Constant curvature vacua of Lovelock theories solve this equation with $\omega=0$.}.  Except in certain special cases, such as Einstein or Gauss-Bonnet gravity 
 or when the Lovelock couplings are tuned such that the polynomial has a unique degenerate root \cite{Crisostomo:2000bb}, the solutions for $f(r)$ are generally not known explicitly.

Here we will use the general relations given in the previous sections to derive expressions for the mass $M$, surface gravity $\kappa$, entropy $S$ and free energy $F$ of Lovelock black holes in terms of $r_h$ without requiring the explicit solution of the field equation.  
This approach provides a geometrical understanding of
  the formulae.
As noted in the introduction, with the application to other even less explicitly known black hole spacetimes in mind, we will take the general form of the metric (\ref{generalmetric}) as our starting point and ``forget" that we know that solutions to the field equations will have $\phi(r)=f(r)$.

\subsection{Constant curvature vacua and Killing potentials}

Working in the Hamiltonian picture,  we assume that for a spherically symmetric static black hole  
the spatial metric has the form
 \begin{equation}\label{spatialmetric}
  s_{ab} dx^a dx^b =  { dr^2 \over f(r) } +r^2 d\Omega ^2 _{D-2},
 \end{equation}
that  the extrinsic curvature $K_{ab}$ vanishes, and that the Hamiltonian
evolution is carried out along the static Killing field $\xi ^a = F n^a$. The function $F^2$ must then have the same
large $r$ fall off conditions as the metric function $\phi$ in (\ref{falloff}) and satisfy $F^2(r_h )=0$. Substituting the spatial metric into the Lovelock Hamiltonian functions
  $H^{(k)}$ in (\ref{hperp}) we find that
we find for these metrics that
   \begin{equation}\label{hkform}
     H^{(k) } = - \gamma _k {1\over  r^{(D-2)} }
{\partial\over \partial r}\left( r^{(D-1-2k)}(1-f)^k \right)
 \end{equation}
where $ \gamma _k ={(D-2)!\over (D-2k-1)! } $.
The constraint equation is
\begin{equation}\label{constreq}
H =\sum _{k=0}^{\kmax} b_k H^{(k)} =0
\end{equation}
First consider the background  AdS metric with radius of  curvature  $l$ and no black hole.
The AdS functions are $F_{AdS}^2 =f_{AdS} =1+r^2 /l^2 $. Since the
metric satisfies the constraint equation with  $f_{AdS}$, substituting 
into (\ref{constreq}) gives a relation 
between the couplings and $l^2$,
 \begin{equation}\label{ldef}
 \sum _{k=0}^{\kmax} (-1)^k {\bhat _k \over l^{2(k-1)} } = 0
  \end{equation}
As noted above, this is  the condition that the sum $s_{(0)}$ defined above in (\ref{sums}) vanish in a constant curvature vacuum.
 Whether, or not, there are real positive solutions for $l^2$ depends on the values of the Lovelock couplings. We will assume
 that we are working with a Lovelock gravity theory that has at least one real, positive solution $l^2$.
 
Next we find the projected Killing-Lovelock potentials $\beta^{(k)rd}n_d $, which are solutions to
equation (\ref{kthhamdiv}) with vanishing shift vector $F^a$. Plugging in the form of the $k$th Hamiltonian (\ref{hkform}) this  becomes
 \begin{equation}\label{tosolve}
 \sqrt{f } {\partial\over \partial r}\left({ r^{(D-2)} \over \sqrt{f}} \beta^{(k)rd}n_d \right) 
= - F \gamma _k
{\partial\over \partial r}\left( r^{(D-1-2k)}(1-f)^k \right) 
\end{equation}
 Now, recall that the existence of the Killing-Lovelock potentials
 depends on the the existence of a
Killing vector $\xi ^a $,  but not on the metric being a solution to the field equations. The statement that the thermodynamic potentials $\Theta^{(k)}$ sum to zero in (\ref{potsum}), however,  does assume that  the constraint equation is satisfied.  Since we will want to make use of (\ref{potsum}) our current task is to solve for the Killing-Lovelock potentials assuming that $f$ solves the constraint 
equation (\ref{constreq}). Note also that  the differential equation (\ref{tosolve}) includes a factor
of the lapse function $F$. In the Hamiltonian picture the lapse  is a gauge choice
 that may be freely specified\footnote{ Note that
 if one  arbitrarily  specifies  $F$ at each subsequent time in the evolution, then
 in general the evolution will not be along a Killing field. To keep the evolution along the Killing
 vector,
 one imposes $\dot{s}_{ab} = \dot{K}_{ab} =0$, which implies a set of differential equations
 that $F$ must satisfy.}.   
 We can choose any function $F$ consistent
 with $-F^2$ being the norm of the static Killing field for a spherically symmetric AdS black hole.
 Specifically,  we can  choose any  $F$ that depends only on $r$ with 
 $F (r_h )=0$ and satisfying the large $r$ fall off conditions in (\ref{falloff}).

  On the other hand, inspection shows 
 that equation (\ref{tosolve}) is easy to solve if we can choose $F = \sqrt{f}$.
 This choice has the correct fall off conditions at large $r$, since the function $f(r)$ does.
 Therefore, we only need to worry about the condition that $F$ vanishes at the horizon, or equivalently whether the function
 $f(r)$ necessarily vanishes at the horizon? Of course it is well known that the 
 Schwarzchild solution in Einstein gravity has $F =\sqrt{f}$, with $f(r)$ vanishing at the horizon.  As noted above
 the static Lovelock black holes continue to have this Schwarzchild form \cite{Wheeler:1985qd}. However, one
 can also demonstrate that  $f(r_h ) =0$ under weaker conditions than requiring that the full set
 of field equations be solved.  Let us assume that
 the surface gravity for the black hole metric (\ref{generalmetric}) is finite and non-zero, and that the constraint equation is satisfied at 
the horizon radius $r_h$. We show in the Appendix that these conditions imply that $f \sim F^2 $ near the horizon, and therefore that
 $f(r_h) =0$.
 Hence we can choose $F=\sqrt{f}$ in (\ref{tosolve}) and the solution for the projected Killing potentials are
\begin{equation}\label{betasol} 
\beta^{(k)rd}n_d = -\gamma _k r^{1-2k} \sqrt{f} (1-f)^k 
\end{equation} 
In the next section we make  use of these solutions to express the mass in terms of the horizon radius $r_h$.

\subsection{Dependence of mass on horizon radius}\label{mass}
It is straightforward to evaluate the boundary integral for the ADM mass (\ref{finalmass})
in terms of the coefficient $c_r$ that characterizes the far field behavior of the spatial metric. With the AdS boundary conditions (\ref{falloff}), one finds
that the terms in (\ref{finalmass}) that  depend
on the derivative of the lapse $F$ cancel, and the integral becomes
\begin{equation}\label{admmasscr}
M= {\Omega _{D-2}s_{(1)} c_r \over 16\pi G  (D-1) } , 
\end{equation}
where we have taken $b_1 =1$.  We next turn to the task of finding the fall-off coefficient $c_r$ and hence also the mass in terms of 
the horizon radius $r_h$ by  applying the results (\ref{theta}) and (\ref{potsum}). The integrand   
 at large $r$ in (\ref{theta}) becomes
  \begin{equation}
  r ^{( D-2k -1)} [ (1-f)^k - (1-f_{AdS})^k ] \simeq (-1)^{k-1}  {k  c_r\over l^{2(k-1)}} ,
  \end{equation}
while the boundary term on the horizon is easily evaluated using the condition $f(r_h )=0$. Assembling these pieces one arrives at the expression for the thermodynamic potentials
\begin{equation}\label{nicetheta}
\Theta ^{(k)} =- \gamma _k \Omega _{D-2} \left(  r_h ^{D-2k -1}  + 
(-1)^k { k c_r \over l^{2(k-1)} } \right)
\end{equation}
where  $\Omega _{D-2}$ is the area of a unit $D-2$ sphere. 
For  $k=0$ this reproduces the result of \cite{Kastor:2009wy} that $\Theta ^{(0)}$ is given by minus an effective volume of the region behind the horizon.
Finally, requiring that the sum rule (\ref{potsum}) be satisfied yields the relation
\begin{equation}\label{crequals}
c_r
= \left( {r_h ^{D-3}\over s_{(1)}}\right) \sum _{k=0}^{\kmax}  { \hat{b}_k \over r_h ^{2(k-1)} }
\end{equation}
This is a key result since it relates the far field behavior of the black hole solution to the horizon radius, without making use of an exact analytic expression for $f(r)$ in the region between.

There are two important points to be made about the expression for $c_r$. First,
we have noted that the solutions for static Lovelock black
holes are known to be specified in terms of solutions to a polynomial equation of order $\kmax$ in the function $f(r)$ \cite{Wheeler:1985qd}.  This equation arises from integrating the Hamiltonian constraint
and $c_r$ is a constant of integration.
 Equation (\ref{crequals}) may also be obtained by evaluating this polynomial at $r_h$ \cite{Cai:2003kt}. 
What we have learned from our more general treatment is that the relation (\ref{crequals}) expresses the fact that
the Hamiltonian constraint is a total divergence when defined with respect to evolution along
a Killing field. Put differently, if a metric does not have a Killing field we have no reason
to expect that the mass is simply a function of data on the horizon, but in general 
will depend on volume integrals as well. 

Second, we can ask whether there a computational advantage to working with the Loveloock potentials and
the sum rule (\ref{potsum}) compared to simply writing out the field equations and analyzing them?
In the spherically symmetric case this is likely a matter of taste.  The analysis in terms of the Killing-Lovelock potentials is more
complicated, but exposes an underlying geometrical structure.
For more complicated spacetimes, such as rotating black holes, it may be that using the Killing potentials 
allows one to find geometrical relations of interest more simply than through a brute force analysis of the field equations.

We can now substitute in for $c_r$ using (\ref{crequals})  and obtain the sought after,
result for the mass of a static Lovelock black hole in terms of its horizon radius
\begin{equation}\label{admmassrh}
M= { \Omega _{D-2} r_h ^{D-3}\over 16\pi G  (D-1)}  \sum _k  { \hat{b}_k \over r_h ^{2(k-1)} }.
\end{equation}
This generalizes to all  Lovelock theories the result for Gauss-Bonnet black holes given in
 \cite{Cai:2001dz}. Such a generalization was assumed to hold in  \cite{Cai:2003kt} and we have now shown that this is indeed the case.

To get an idea of how this formula behaves, let us examine the dependence of the mass on $r_h$ in different limiting regimes.
The behavior of the mass as $r_h \rightarrow 0$ turns out to be
interesting in that it differs between even and odd spacetime dimensions. 
Since the small black holes are dominated by the
 highest curvature terms, behavior in this limit depends on whether the order $\kmax$ term actually appears in the Lagrangian with non-zero coupling constant.   We will assume that this is the case.
Note that this includes Gauss-Bonnet gravity in $D=5$ and in $D=6$, for which $\kmax=2$, but not in higher dimension.

By a  ``small"  black hole will be one such that the horizon radius is sufficiently small that 
$r_h^{2(\kmax-k)}\ll |\hat b_{\kmax}/\hat  b_k|$ for all $k<\kmax$.
After noting  that  $\kmax =(D-1)/2$, and $2\kmax =(D-2)/2$ respectively in odd and even dimensions, one finds that in the small black hole regime the mass depends on $r_h$ as
\begin{eqnarray}
&M\approx  {\Omega _{D-2}\over 16\pi G (D-1)}\bhat _{\kmax} r_h\qquad & D\,\, even\\
&M\approx  {\Omega _{D-2}\over 16\pi G (D-1)}\left( \bhat _{\kmax} + \bhat _{{\kmax} -1 }r_h ^2
\right)\qquad & D\,\, odd
\end{eqnarray}
However, for $D$ odd the mass goes to a nonzero value,
We see that for even dimensions, the mass goes smoothly to zero with the horizon radius as it does for Schwarzschild black holes.
However, for odd dimensions there is a minimum mass for black holes that is proportional to $\hat b_{\kmax}$.
This minimum mass, or mass gap, has been discussed previously for $D=5$ Gauss-Bonnet black holes in \cite{Cai:2001dz} and for Chern-Simons-Lovelock theories, which have a unique constant curvature vacuum, in \cite{Crisostomo:2000bb}.

So long as the coefficient $b_0$ of the cosmological constant term in the Lovelock Lagrangian is non-zero,
black holes in the opposite regime of large $r_h$ are dominated by the cosmological constant and
look qualitatively the same in all dimensions. A``large" large black hole will be one such that the horizon radius satisfies 
$r_h^{2k}\gg |\hat b_k/\hat b_0|$ for all $k>0$.
For large black holes one finds that the mass depends on the horizon radius as
\begin{equation}\label{largemass}
M\approx   {\Omega _{D-2}\over 16\pi G (D-1)}b_0 r_h ^{D-1} .
\end{equation}

\subsection{Surface gravity of Lovelock black holes}\label{surfacegrav}

For a metric of the general form (\ref{generalmetric}) the surface gravity is given by $\kappa =(1/2) \phi ' (r_h )$.
In the appendix we argue  that  near the horizon of a black hole the metric functions must
satisfy $g_{tt} \approx - 1/g_{rr} $, with 
 equality  for the functions and their first derivatives at the horizon. 
Hence $f(r_h )=0$ and $ f' (r_h )= 2\kappa  $.  Evaluating the
Hamiltonian function $\hk$ in (\ref{hkform}) at the horizon and  applying the sum rule (\ref{potsum})
yields a general relation between the surface gravity and the horizon radius without the need for the explicit form of the metric function $f(r)$,
\begin{equation}\label{surfgrav}
\kappa ( \sum _{k=0}^{\kmax}    {k\,\hat{b}_k \over r_h ^{2(k-1)} } ) =
{1\over 2 r_h } \sum _{k=0}^{\kmax}  { (D-2k-1)\hat{b}_k \over r_h ^{2(k -1)}}
\end{equation}
For Gauss-Bonnet gravity this agrees with the expressions for the surface gravity obtained from the explicit solutions in \cite{Cai:2001dz,Cvetic:2001bk,Cai:2003kt} using 
an approach similar to the one here.

Let us examine at the behavior of the surface gravity in various limits regimes starting with small black holes as defined above.
In this regime, one finds that
\begin{eqnarray}\label{smallkappaone}
 & \kappa  \approx {1\over 2\kmax\, r_h } ,
\qquad & D\  even \\  \label{smallkappatwo}
 & \kappa \approx  {\bhat _{\kmax}r_h \over \kmax\bhat _{({\kmax }-1)} }
   ,\qquad &D\  odd 
  \end{eqnarray}
Again, we see a qualitative difference between even and odd dimensions.
In even dimensions the surface gravity diverges in the limit of vanishing horizon radius, as it does for Schwarzschild black holes in $D=4$.
However for generic Lovelock theories in odd dimensions, {\it i.e.} those in which $b_{\kmax}$ is nonvanishing, the surface gravity goes smoothly to zero with the horizon radius.
This change in behavior was noted for Gauss Bonnet gravity in $D=5,6$ in \cite{Cai:2001dz}.
Here we note, in agreement with the observations of   \cite{Cai:2003kt}, that this is characteristic of Lovelock black holes in general.

 On the other hand, the surface gravity of large black holes 
 has qualitatively the same behavior for all AdS-Lovelock  black holes.  One finds that
\begin{equation}\label{largekappa}
\kappa \approx {b_0 r_h \over 2(D-2)} 
\end{equation}
where we have set $b_1 =1$, so that $\bhat _1 =(D-1)(D-2)$.
In any dimension, therefore, there exists a large black hole solution for sufficiently high
temperature.  In even dimensions also always exists a small black hole at suficiently  high temperatures.
Hence for $D$ even, and so long as the Lovelock couplings are such that the surface gravity stays positive for all horizon radii, there will be 
a minimum temperature at which static black holes exist, while  in 
odd dimensions there  is no minimum temperature. In addition, there may be local extrema of the temperature depending
 on the choices of the $b_k$.

 \subsection{Entropy, thermodynamic potential and Smarr relation}

The expression for the entropy \cite{Jacobson:1993xs} given in section (\ref{kl_potentials}) is a sum of integrals  of Lovelock invariants constructed from the induced metric $\gamma_{ab}$ on the horizon cross section.  For a spherically symmetric black hole $\gamma _{ab}$ is the metric of a round $D-2$ dimensional sphere of radius $r_h$ and the Riemann tensor is given simply by $R_{ab}{}^{cd}=(1/r_h^2)\, \delta_{ab}^{cd}$.  Evaluating the various Lovelock terms explicitly gives
$\call^{(k-1)} (\gamma_{ab}) =  {(D-2)! \over (D-2k)! }\, r_h ^{-2(k-1)}$.
The entropy for a static, spherically symmetric Lovelock black hole can then be written as
\begin{equation}\label{sphentropy}
S
= {\Omega _{D-2} r_h ^{D-2} \over 4(D-1)G}\,\,
 \sum_{k=0}^{\kmax} {k\, \bhat_k  \over (D-2k)\, r_h ^{2(k-1)}}.
\end{equation}
We can also compute the quantity $S^\prime$ defined in (\ref{extras}) which appears in the Smarr formula.  It is given in terms of the horizon radius by
\begin{equation}\label{sprime}
S^\prime
= {\Omega _{D-2} r_h ^{D-2} \over 4(D-1)G}\,\,
 \sum_{k=0}^{\kmax} {2k(k-1)\, \bhat_k  \over (D-2k)\, r_h ^{2(k-1)}}.
\end{equation}

Our goal in this section has been to develop expressions for the thermodynamic properties of AdS-Lovelock black holes purely in terms of the horizon radius.  The final element of the Smarr relation (\ref{smarr}) is the overall thermodynamic potential $\Theta$, which we will now compute.
With this in hand, we can check all of our results are consistent with the Smarr relation.
Substituting our result (\ref{crequals}) for $c_r$ into
the expression (\ref{nicetheta}) for the $\theta^{(k)}$ and performing the sum over $k$ in (\ref {extras}) gives the result
\begin{equation}\label{sumthetatwo}
\Theta =  - {\Omega _{D-2} r_h ^{D-3}\over 8\pi G (D-1)} \sum_{k=0}^{\kmax}\left( k -1 +   {s_{(2)} \over s_{(1)} }   \right)
    { \hat{b}_k \over  r_h ^{2(k-1)} }
 \end{equation}
Combining the results for the mass (\ref{admmassrh}), surface gravity (\ref{surfgrav}), entropy (\ref{sphentropy}), $S^\prime$ in (\ref{sprime}) and thermodynamic potential (\ref{sumthetatwo}), it is now straightforward to check that the validity of the Smarr formula (\ref{smarr}) for the AdS-Lovelock black holes.

\section{Free energy and phase transitions}\label{freesection}

In this section, we will make use of our results to give a general expression for the free energy of 
AdS-Lovelock black holes.  We will discuss the behavior of this expression in the limits of large and small horizon radius, in which it is independent of the detailed choice of Lovelock couplings.

The free energy of asymptotically AdS black holes has been a topic of interest since the work of Hawking and Page \cite{Hawking:1982dh} in Einstein gravity.  They found that there exists a minimum temperature $T_0$ for such black holes which occurs for a horizon radius $r_0$.  The temperature diverges both in the limit of large black holes and in the limit of zero horizon radius.  Black holes with horizon $r_h<r_0$, like asymptotically flat black holes, have negative specific heat and cannot be in stable equilibrium with a thermal bath of radiation.  However, solutions with $r_h>r_0$ have positive specific heat and can be in stable equilibrium.  For large black holes with temperatures just above $T_0$, the free energy is positive and thermal AdS space, with zero free energy, represents that globally preferred thermodynamic state.  However, the free energy of large black holes becomes negative above a critical temperature $T_1>T_0$ (or correspondingly for black holes with radii exceeding a certain threshold $r_1$) and the black hole is then the globally preferred state.

Hawking and Page found the free energy by computing the Euclidean action, which requires the analytic form of the AdS-Schwarzschild spacetimes.  Similar computations of the free energy have been carried out in Gauss-Bonnet gravity \cite{Cai:2001dz,Cvetic:2001bk,Nojiri:2001aj,Cho:2002hq}, 
where the explicit solutions are also known \cite{Boulware:1985wk,Wheeler:1985nh}.
An expresssion for free energy in general Lovelock gravity is given in \cite{Maeda:2011ii} which is calculated using a generalized quasi-local mass defined for spherically (plane or hyperbolic) symmetric spacetimes. 
However, theses computations are considerably more complicated.  Our method offer a much a simpler route to the result  and  offers some degree of physical interpretation to the different terms in the result.
It also yields the answer in the general Lovelock case, where do to the absence of explicit analytic solutions, computation of the Euclidean action may not be practical.  Moreover as stated above, we envision further applications to even less well understood solutions, such as rotating Lovelock black holes.

It is worth noting explicitly that the equality of the free energy with the Euclidean action times the temperature continues to hold generally in Lovelock gravity theories.
Hawking and Horowitz  \cite{Hawking:1995fd} used  the Hamiltonian framework for Einstein gravity to demonstrate that
 $I_E =\beta M -S$ where $\beta$ is the Euclidean period. 
A similar construction works in Lovelock gravity, the basic steps being as follows.

Write the volume term of the Euclidean action for a static black hole in Hamiltonian variables,
and then directly derive the boundary term for the action by varying this expression. 
The variation of the volume term is the same calculation that is done to find the mass (see \cite{Crisostomo:2000bb,Kastor:2010gq}) with an additional integration over Euclidean time.  Evaluated on static solutions, the volume 
term of the action reduces to a sum of the constraints and hence vanishes. 
The value of the action is then given by the boundary terms.  At infinity the  boundary
term is simply $M/T$.   The Euclidean metric does not have a horizon.  However, Hamiltonian
evolution with respect to Euclidean time fails to be well defined at $r_h$.  In order to compensate for this, one introduces an inner boundary at $r=r_h +\epsilon$. In the limit
$\epsilon \rightarrow 0$ the boundary term is equal to the entropy $S$. 
One then has the result that the free energy defined by $F=TI_E$ coincides with the thermodynamic free energy $F=M-TS$.
Note also that the free energy has been defined such that it vanishes for 
AdS\footnote{However, because of the presence of a mass gap for odd 
dimensional AdS-Lovelock black holes, the limit $r_h\rightarrow 0$ will not vanish in our 
subsequent expressions for  $F$ given below. }.

Let us see how the Hawking-Page transition arises in the present framework.  In Lovelock gravity the cosmological constant is given by 
$\Lambda=-b_0/2$, so that $b_0$ is positive for $\Lambda$ negative.
Starting from $F=M-TS$, let us use the Smarr formula (\ref{smarr}) 
to eliminate the mass.  For Einstein gravity the quantities  $s_{(2)}$ and $S^\prime=0$ in the Smarr relation vanish, while as shown in \cite{Kastor:2009wy} the overall Lovelock thermodynamic potential is given by 
$\Theta = - V_{bh} $
where $V_{bh}$ is an effective volume for the black hole given by the flat (or AdS) spacetime volume of a sphere of radius $r_h$.  One then arrives at an expression for the free energy
 \begin{eqnarray}\label{hpfree}
 F&=&{1\over 8\pi G(D-3)} \left(\kappa A -b_0 V_{bh}\right)\\
    &=& { \Omega _{D-2} r_h^{D-2} \over 8\pi G  (D-3)  } \left( \kappa - { b_0   r_h \over (D-1) } \right) 
\end{eqnarray}
The important observation is that the apparent contribution of $\kappa A$ to the free energy has changed sign in (\ref{hpfree}) relative to the original expression $F=M-TS$.
This is because  the mass
receives a positive contribution from $\kappa A$ in the Smarr formula and the overall net coefficient is always positive. For an asymptotically flat black hole, {\it i.e.} with vanishing $b_0$, this is the
only term and the free energy is always positive. With a negative cosmological constant there is
 a negative definite contribution that takes the form of a cosmological pressure $b_0$ times an
 effective volume of the black hole $V_{bh}$. Hence the Hawking-Page phase transition at which the free energy changes sign may be thought of as arising from the $\Theta$ term in the Smarr relation, which is itself analogous to
a $PV$-type contribution in classical thermodynamics.

 To determine whether
the positive or negative term dominates $F$
one needs to use know how the surface gravity behaves. Hawking and Page used the
analytic solutions to compute $\kappa$ and found that
the free energy is positive for small black holes as in Schwarzchild. However,
they found that large black holes have negative free energy.  Indeed, substituting
$\kappa$ from equation (\ref{surfgrav}) one recovers these results. However, rather than recall this case
in more detail, we turn to an analysis of the free energy for general Lovelock black holes, 
which includes the case of Einstein gravity. We will see that  the behavior
of the free energy for large black holes is qualitatively the same as the Hawking-Page case. On the 
other hand, the free energy of small black holes differs between even and odd
dimensions due to the alternating behavior of the surface gravity.

\subsection{Thermodynamic stability and phase transitions for Lovelock black holes}

There are a number of contributions to the general Lovelock free energy and the simplification made above in the Einstein case using the Smarr formula no longer yields an easily interpretable expression.  Instead we simply substitute into the free energy the formulas for the mass  (\ref{admmassrh}) and the entropy (\ref{sphentropy}) to obtain
\begin{equation}\label{morefree}
 F  =  { \Omega _{D-2} r_h^{D-3} \over 16\pi G   (D-1)  }
   \sum_{k=0}^{\kmax} {\bhat_k \over r_h ^{2(k-1)} } 
  \left( 1-    \kappa  r_h   { 2k \over (D-2k)} \right) 
\end{equation}
where $\kappa$ can also be regarded as in (\ref{surfgrav}) as a function of the horizon radius and the Lovelock couplings.
This expression for $F$ agrees with that of reference \cite{Cai:2001dz}
in the case of Gauss-Bonnet gravity and \cite{Maeda:2011ii} in general Lovelock gravity, but disagrees\footnote{
The expression for the free energy in \cite{Cvetic:2001bk}
is obtained by computing the volume term in the  Euclidean action,
apparently without the inclusion of a boundary term. A subtraction of  the action for pure $AdS$ at large radius is used to regularize 
the result.  There are then several differences
with our calculation.  First, we implicitly use the Euclidean action with
the boundary term that gives a well defined variational principal in the Hamiltonian variables. 
As described above the action on solutions is then given entirely by the boundary term and is equal to the finite quanity $M-TS$, without a
need for regularization.}
 with that given in reference \cite{Cvetic:2001bk}.
 
Let us start by examining the behavior of the free energy in the large
black hole limit. Form (\ref{largekappa}), we see that the surface gravity grows  like $b_0\, r_h$ for large $r_h$ in all dimensions.
One then finds that  there are both positive and negative contributions at leading order growing like $b_0 r_h^{D-1}$.
The net result for the free energy in the large black hole limit turns out to be negative,
\begin{equation}\label{largef}
F \approx -{ \Omega _{D-2} b_0 r_h ^{D-1} \over 16\pi G  (D-1)(D-2)  } 
\end{equation}
This result was also pointed out in \cite{Camanho:2011rj} (see also 
\cite{Cai:2006pq,Cai:2009de}) and  is not surprising since the behavior in this regime is 
dominated by the cosmological constant and Einstein terms. 

The behavior of the free energy (\ref{morefree}) in the small black hole regime, on the other hand, is dominated by the highest curvature
terms  and differs between even and odd dimensions. In even dimensions both the mass $M$ and the product $\kappa S$ 
 scale like $r_h$ in the limit of small horizon radius.
 One finds in this case that  the positive contribution to the free energy coming from the mass dominates, giving
\begin{equation}\label{smallevenf}
F \approx { \Omega _{D-2}\,\hat b_{\kmax} r_h\over 32\pi G  (D-1)  }  , \qquad 
 D\,\,  even
\end{equation}
In odd dimensions both the surface gravity and entropy vanish as $r_h\rightarrow 0$ (see equation (\ref{smallkappatwo})),
while the mass has a finite positive limiting value, giving
\begin{equation}\label{smalloddf}
F\approx { \Omega _{D-2} \,\hat b_{\kmax} r_h\over 16\pi G  (D-1)  } , \qquad    D\,\,  odd
\end{equation}
Hence $F$ is positive for small black holes in all dimensions. For $D$ even
$F$ goes to zero, while for $D$ odd $F$  goes to a nonzero positive value.

It is also straightforward to compute the specific heat
in the large and small black hole limits.
Using the expressions  for the mass and the surface
gravity in section (\ref{mainsection}) we find for large black holes
\begin{equation}\label{heatlarge}
{\partial M \over \partial T} \approx {(D-2)\Omega _{D-2}\,  r_h ^{D-2}  \over 4G}  
\end{equation}
and for small black holes
\begin{eqnarray}\label{heatsmalleven}
{\partial M\over  \partial T} &\approx -{ \kmax\,\Omega _{D-2}\, {\hat b}_{\kmax} r_h ^2\over 4G (D-1)} 
\qquad  &D\,\, even\\
\label{heatsmallodd}
{\partial M\over  \partial T}& \approx {\kmax\, \Omega _{D-2}\, {\hat b}^2 _{(\kmax - 1)} \, r_h \over 4G (D-1){\hat b}_{\kmax}} 
\qquad &D\,\,  odd
\end{eqnarray}
where we have assumed that the couplings $b_0$,  $b_{\kmax}$ and $b_{\kmax -1}$ are all nonzero.  

Let us now summarize these results. While a detailed understanding of the global and local thermodynamic stability of AdS-Lovelock black holes throughout the entire range of horizon radii would require specifying the entire set of Lovelock couplings, the behavior in the large and small black hole regimes is generic.  In dimensions, we find that the behavior of AdS-Lovelock black holes in these regimes is similar to that found by Hawking and Page in Einstein gravity  \cite{Hawking:1982dh}.  Black holes become arbitrarily hot in both limits.  Small black holes in even dimensions exhibit negative specific heat and have positive free energy, indicating instability to both perturbative and non-perturbative fluctuations.
Large black holes, on the other hand, 
have positive specific heat and negative free energy, both indicating their stability to small thermal fluctuations
and that they are  the thermodynamically globally preferred state.   If one assumes that the Lovelock couplings are such that the temperature stays positive for all $r_h$, then there must exist a minimum temperature $T_{0}$ below which no black hole solutions exist.  As a consequence of these similarities, we can expect that even dimensional Lovelock black holes will at least have a simple Hawking-Page phase transition and possibly a more complicated structure of phase transitions, depending on the detailed behavior of the temperature and free energy over the whole range of horizon radii.
 
In odd dimensions only  large black holes
exist at very high  temperatures. They  have positive specific heat and negative free energy
as in even dimensions and are therefore thermodynamically stable. 
The odd dimensional  small black holes, on the other hand, have positive positive free energy and
also have positive specific heat. So they are stable to small, but not large, thermal fluctuations, and unlike small 
black holes is even dimensions can be in stable equilibrium with a thermal bath.

In odd dimensions the low temperature picture is different. This was studied 
in $D=5$ Gauss-Bonnet gravity in references \cite{Cai:2001dz,Cho:2002hq} .
 Black holes exist for \cite{Cai:2001dz} temperatures down to $T=0$, and the low temperature black holes have positive
specific heat, so there is a locally stable alternative to the gas state. At these low
 temperatures  the pure gas state has lower free energy since $F_{gas}\rightarrow 0$
 as $T\rightarrow 0$, while the black hole starts with $F$ of order $b_{\kmax} $. 
 Still, a small black hole can exist in equilibrium with a low temperature gas, unlike the
 situation in even dimensions. So the behavior of small black holes in $D=5$ Gauss-Bonnet
 gravity continues in odd dimensions as long as the highest curvature Lovelock term is
 included. For very high temperatures there is only the one  large black hole state. As
 in even dimensions this is both globally and locally preferred.

\section{Discussion}\label{discussion}
In this paper we started by deriving  a fully explicit formula
for the ADM mass of an asymptotically AdS spacetime
in a generic Lovelock gravity  theory,  via the Hamiltonian methods
 of Regge \& Teitelboim  \cite{Regge:1974zd}. We then proceeded to study various thermodynamic properties of AdS-Lovelock black
holes.  In particular, we made use of the Killing-Lovelock  potentials that exist
in these spacetimes in order to evaluate the mass in terms of the horizon radius
and the Lovelock couplings. After finding expressions for the surface gravity and
entropy, these ingredients were assembled to give the free energy. All of these expressions
are quite general, assuming only that solutions exist with the prescribed asymptotic forms.

As mentioned in the introduction, we envision further applications of these techniques to stationary solutions that are even ``more unknown", such as higher dimensional rotating charged black holes, or rotating Lovelock black holes.  Another possible direction for future
work would be to look at black holes/branes in AdS with planar horizons.   In this case 
the asymptotic boundary of a spatial slice is a plane rather than a sphere.
If one  of these directions is compact with length $L$, then it would be necessary to further
extend the first law to include an appropriate $\delta L$ term as in the asymptotically 
flat Kaluza-Klein case \cite{Kastor:2006ti}. 

\subsection*{Acknowledgements}

The work of DK and JT  was supported by NSF grant PHY-0555304.  
DK and JT  also acknowledge the hospitality of the Centro de Ciencias de Benasque, Spain where this
work was begun.
The work of SR was funded by FONDECYT grant 3095018 and by the CONICYT
grant \textquotedblleft Southern Theoretical Physics Laboratory
\textquotedblright\ ACT-91. Centro de Estudios Cient\'{\i}ficos (CECS)
is funded by the Chilean Government through the Millennium Science
Initiative and the Centers of Excellence Base Financing Program of
CONICYT. CECS is also supported by a group of private companies which at
present includes Antofagasta Minerals, Arauco, Empresas CMPC, Indura,
Naviera Ultragas and Telef\'{o}nica del Sur. CIN is funded by CONICYT
and the Gobierno Regional de Los R\'{\i}os.

\renewcommand{\theequation}{A\arabic{equation}}
  \setcounter{equation}{0}  
  \section*{Appendix - Near horizon behavior of  $g_{rr}$}  

The horizon of a static black hole occurs where the norm of the static Killing field is zero.
In the coordinates of the metric (\ref{generalmetric}) this is at $r_h$ such that $\phi (r_h )=0$.
In this appendix we show that if one assumes that 
 the surface gravity  is finite and that  the Hamiltonian constraint equation  is satisfied 
 at $r_h$, then near the horizon it follows that
  $f(r)\sim \phi (r )$, and in particular that $f(r_h )=0 $ , $f' (r_h )= 2\kappa $.
  
 We point out that for the spherically symmetric black hole metric (\ref{generalmetric})  
it has been shown \cite{Wheeler:1985nh} \cite{Wheeler:1985qd}     that the vacuum field
equations imply that $- g_{tt}= 1/g_{rr}$, or $\phi (r) =f(r) $ everywhere. Clearly this is a  stronger result
than the result shown here. 
However, to derive the expressions for $M$, $\kappa$, and the $\Theta ^{(k)}$ we only
 need to use the values of $f$ and $f'$ at $r_h$, which can be found with correspondingly less work,
 as follows.

Assume that the metric function $\phi$ in (\ref{generalmetric}) goes
to zero like a power law as $r$ approaches $r_h$, {\it i.e.} that near the horizon
\begin{equation}\label{phih}
\phi \simeq \phi _1 (r-r_h ) ^ p
\end{equation}
where $\phi_1$ is a constant.
The surface gravity, given by $\kappa ^2 =-{1\over 2}( \nabla _a \xi _b) \nabla ^a \xi ^ b$,
then becomes
\begin{equation}\label{sg}
\kappa ^2 = 
 {1\over 2} p^2 \lim_{r\rightarrow r_h}(r-r _H )^{p-2} f(r)
\end{equation}
In order for $\kappa^2$ to be finite, it must be that the metric function $f(r)$ behaves  like
\begin{equation}\label{fh}
f \simeq f _1 (r-r_h ) ^ {2-p} 
\end{equation}
with $f_1$ another constant.  Let us rewrite the expression (\ref{hkform}) for  $H^{(k)}_\perp$
as
\begin{equation}
H ^{(k)} = r^{1-2k} \left[(D-2k-1)(1-f)/r -k(1-f)^k f'  \right]
\end{equation}
and consider the constraint equation (\ref{constreq}).
If the power law index $p>1$, we see from line (\ref{fh}) that $f' $ diverges at the horizon, and inspection shows that 
$H=0$ cannot be satisfied at $r_h$. On the other hand, if the power law index $p<1$, then $f(r_h ) = f'(r_h ) =0$ and again the
constraint cannot be satisfied\footnote{If the spacetime is  not vaccuum, the the gravitational
constraint is $H=-16\pi\rho$, and so powers $p<1$ are not ruled out.} at $r_h$.
Hence, in order to have  finite, non-zero surface gravity, the power law index in (\ref{sg}) and (\ref{fh}) must be $p=1$, so that 
$\phi \sim \phi _1 (r-r_h )$  and $f \sim f_1 (r-r_h ) $, which gives $\kappa ^2  ={1\over 2} \phi _1 f_1$.
We see that the overall numerical scale of $\kappa $ is not fixed by this argument,
which makes sense because the normalization of $\kappa$ is fixed by the norm of the time-translation Killing vector at infinity, which requires knowing the function 
 $\phi (r)$ throughout the spacetime. However, one can fix the scale
 by requiring that it gives the right answer for Schwarzchild.

\end{document}